\documentclass[10pt,a4paper]{article}

\usepackage{amsmath}
\usepackage{amsthm}
\usepackage{amssymb}
\usepackage{enumerate}

\newtheorem{theorem}{Theorem}[section]
\newtheorem{lemma}[theorem]{Lemma}
\newtheorem{definition}[theorem]{Definition}

\newtheorem{pppp}{Proof}

\newenvironment{pf}{\begin{pppp} \em}{\mbox{}\hfill\qed\end{pppp}}

\newtheorem{FFF}{Remark}

\newtheorem{GGG}{Note}

\newtheorem{HHH}{Problem}

\newtheorem{AAA}{Input : }

\newtheorem{BBB}{Output : }

\renewcommand{\appendix}{%
    \renewcommand{\section}{%
        \newpage\thispagestyle{plain}%
        \secdef\Appendix\sAppendix}%
    \setcounter{section}{0}%
    \renewcommand{\thesection}{\Alph{section}}%
}

\begin{document}

\begin{center}{\Large\bf Construction of optimal codes in deletion and insertion metric}\\
[15pt] Hyun Kwang Kim\footnotemark[1], Joon Yop
Lee\footnotemark[2], and Dong Yeol Oh\footnotemark[1]
\end{center}
 \footnotetext[1]{Department of
Mathematics, Pohang University of Science and Technology, Pohang
790-784, The Republic of Korea (hkkim@postech.ac.kr, \
dyoh@postech.ac.kr).}
 \footnotetext[2]{ASARC, KAIST, Daejeon
305-340, Korea (flutelee@postech.ac.kr).}


\begin{abstract}
We improve Levenshtein's upper bound  for the cardinality of a
code of length four that is capable of correcting single deletions
over an alphabet of even size. We also illustrate that the new
upper bound is sharp. Furthermore we construct an optimal perfect
code that is capable of correcting single deletions for the same
parameters.
\end{abstract}

\section{Introduction}

Let $B_q=\{0,\,1,\,\ldots,\,q-1\}$ be a set with $q$ elements.
$B_q$ is referred to as an alphabet and its elements are referred
to as letters. A sequence $x=(x_1,\ldots,x_n)$ of $n$ letters of
$B_q$ is called a word, and the number $n$ is called its length.
Together with writing $x=(x_1,\ldots,x_n)$, we will also use the
notation $x=x_1\cdots x_n$. Let $B_q^n$ be the set of words over
$B_q$ of length $n$, and define
$$B_q^*=\bigcup_{n=0}^\infty B_q^n.$$

The deletion and insertion distance $\rho(x,y)$, which was first
introduced by Levenshtein \cite{L1}, between two words $x$ and $y$
in $B_q^*$ is defined by the minimum number of deletions and
insertions of letters required to transform $x$ into $y$. For
example, let $x=12243$ and $y=14223$ be two words in $B_5^5$.
Deleting the fourth letter of $x$ and the second of $y$ yields the
identical words $x'=1223$ and $y'=1223$, respectively. Hence
$\rho(x,y)=2$. For a code $C \subseteq B_q^n$ with $|C|\ge 2$, we
define
$$\rho(C)=\min\left\{\ \rho(x,y)\ |\ x,y\in C,x\neq y\right\}.$$

Let $N(n,q,d)=\max\left\{\ |C|\ |\ C \subseteq B_q^n, \rho(C)>2d
\right\}.$ A code $C$ in $B_q^n$ is called an optimal code if
$|C|=N(n,q,d)$. Levenshtein \cite{L} estimated an upper bound of
$N(n,q,1)$ for any $n\ge 2$ and $q\ge 2$, namely,
\begin{equation}\label{bound0}
N(n,q,1)\leq
\left\lfloor\frac{q^{n-1}+(n-2)q^{n-2}+q}{n}\right\rfloor.
\end{equation}
The upper bound in (1) is sharp when $n=3$. Indeed, a code that is
capable of correcting single deletions whose cardinality meets the
upper bound in (1) was constructed in \cite{L} when $n=3$ and
$q\ge 2$ is an arbitrary integer. However this bound is not sharp
when $n \geq 4$. In this paper, we improve Levenshtein's upper
bound for the case in which $n=4$ and $q$ is even, and prove that
the new upper bound is sharp.

In Section 2, we will derive a new upper bound for the cardinality
of a code that is capable of correcting single deletions when
$n=4$ and $q$ is even by analyzing the deletion map.

In Section 3, we will construct a code in $B_q^{4}$ for $q \equiv
2\ \text{or}\ 4\ (\text{mod}\ 6)$ that is capable of correcting
single deletions whose cardinality meets the new upper bound
derived in Section 2. We first introduce the concept of the step
property for a Steiner quadruple system, and prove that there is a
code in $B_q^{4}$ that is capable of correcting single deletions
whose cardinality meets the upper bound derived in Section 2,
under the assumption that there is a Steiner quadruple system on
$B_q$ that satisfies the step property. Then we show that there is
a Steiner quadruple system on $B_q$ that satisfies the step
property for $q \equiv 2\ \text{or}\ 4\ (\text{mod}\ 6)$.

In Section 4, we will construct a code in $B_q^{4}$ for $q \equiv
0\ (\text{mod}\ 6)$ that is capable of correcting single deletions
whose cardinality meets the upper bound derived in Section 2.
Since there does not exist a Steiner quadruple system on an
alphabet of this size, we will use a group divisible system. We
divide our construction into two steps. In the first step, we
prove that an optimal code exists for an alphabet of size $q = 6m$
where m is odd. In the next step, we prove that an optimal code
exists for an alphabet of size $2q$ under the assumption that an
optimal code exists for an alphabet of size $q$.

In Section 5, we modify our construction of optimal codes
slightly, and construct an optimal perfect code in $B_q^4$ when
$q$ is even.

\section{New upper bound of $N(4,q,1)$ for $q$ even \label{bound}}

In this section, we improve Levenshtein's upper bound for the case
in which $n=4$ and $q$ is even. Our result follows from an
analysis of the deletion map as follows.

We begin with a simple observation. For any word $x$ over $B_q$
and any positive integer $s$, denote by $\lfloor x\rfloor_s$ the
set of words obtained from $x$ by deleting $s$ of its letters. The
multi-map $\left\lfloor\  \right\rfloor_s: B_q^n \rightarrow
B_q^{n-s}$ which sends $x$ to $\left\lfloor x\right\rfloor_s$ will
be called an $s$-deletion map or simply a deletion map.  For a
subset $C\subseteq B_q^n$, denote by $\lfloor C \rfloor_s$ the set
$\bigcup \limits_{x\in C}\left\lfloor x\right\rfloor_s$. A set $C
\subseteq B_q^n$ is called a code that is capable of correcting
$s$ deletions if all sets $\lfloor x\rfloor_s\ \left(x\in
C\right)$ do not pairwise intersect. It follows from the
definition that a code $C$ in $B_q^n$ satisfies $\rho(C)>2s$ if
and only if $C$ is capable of correcting $s$ deletions.

The following lemma is an immediate consequence of this
observation.
\begin{lemma}[Substitution lemma] Let $C$ be a code in $B_q^n$
such that $\rho(C)>2s$. If $x\in C$, $y\in B_q^n$, and
$\left\lfloor y\right\rfloor_s\subseteq \left\lfloor
x\right\rfloor_s$, then $\rho((C\backslash\{x\})\cup\{y\})>2s$.
More generally, if $A\subseteq C$, $B\subseteq B_q^n$,
$\left\lfloor B\right\rfloor_s\subseteq \left\lfloor
A\right\rfloor_s$, and $\rho(B)>2s$, then $\rho((C\setminus A)\cup
B)>2s$.
\end{lemma}

From now on, we will use $C$ to denote a code in $B_q^4$ with
$\rho(C)>2$ and we define
$$C_i=\{x\in C \bigm | |\left\lfloor x \right\rfloor_1 |=i\}\ (1\leq i\leq 4).$$
Let $a$, $b$, $c$, and $d$ be distinct elements in $B_q$. By the
substitution lemma, we may assume without changing the cardinality
of $C$ that
$$\left\{\begin{array}{l}
x\in C_1 \Rightarrow x\ \text{is of the type}\ (a,a,a,a),\\
x\in C_2 \Rightarrow x\ \text{is of the type}\ (a,a,b,b),\\
x\in C_3 \Rightarrow x\ \text{is of the type}\
(a,b,b,a),(a,a,b,c),(b,c,a,a),\ \text{or}\ (a,b,b,c),\\
x\in C_4 \Rightarrow x\ \text{is of the type}\
(a,b,c,d),(a,b,a,c),(a,b,c,a),\ \text{or}\ (b,a,c,a).
\end{array}\right.$$ In the case of $3\leq i\leq 4$, we analyze further and
define
$$\left\{\begin{array}{l}
C_{3,1}=\{x\in C_3 \bigm | x\ \text{is of the type}\
(a,a,b,c),(b,c,a,a),\
\text{or}\ (a,b,b,c)\},\\
C_{3,2}=\{x\in C_3 \bigm | x\ \text{is of the type}\ (a,b,b,a)\},\\
C_{4,1}=\{x\in C_4 \bigm | x\ \text{is of the type}\ (a,b,c,d)\},\\
C_{4,2}=\{x\in C_4 \bigm | x\ \text{is of the type}\ (a,b,c,a)\},\\
C_{4,3}=\{x\in C_4 \bigm | x\ \text{is of the type}\ (a,b,a,c)\
\text{or}\ (b,a,c,a)\}.\end{array}\right.$$ Consider the following
subsets of $B_q^3$:
$$\left\{\begin{array}{l}
U=\{y\in B_q^3 \bigm | y\ \text{is of the type}\ (a,a,a)\},\\
V=\{y\in B_q^3 \bigm | y\ \text{is of the type}\ (a,a,b)\
\text{or}\
(a,b,b)\},\\
W=\{y\in B_q^3 \bigm | y\ \text{is of the type}\ (a,b,a)\},\\
Z=\{y\in B_q^3 \bigm | y\ \text{is of the type}\
(a,b,c)\}.\end{array}\right.$$ Note that
\begin{eqnarray}\left\{\begin{array}{l}
\left|U\right|=q, \ \left|V\right|=2q(q-1),\\
\left|W\right|=q(q-1), \ \left|Z\right|=q(q-1)(q-2).
\end{array}\right.
\end{eqnarray}
Table 1 counts the contribution of each codeword in $C_i$ (or
$C_{i,j}$) to $U$, $V$, $W$, and $Z$ under the deletion map
$\left\lfloor\ \right\rfloor_1:B_q^4\rightarrow B_q^3$.
\begin{table}[h]
  \centering
\begin{tabular}{|c|c|c|c|c|c|c|c|}\hline
   &$C_1$&$C_2$&$C_{3,1}$&$C_{3,2}$&$C_{4,1}$&$C_{4,2}$&$C_{4,3}$\\\hline
$U$&1&0&0&0&0&0&0\\ \hline $V$&0&2&2&2&0&0&1\\\hline
$W$&0&0&0&1&0&2&1\\ \hline $Z$&0&0&1&0&4&2&2\\\hline
\end{tabular}
\caption{The contribution of each codeword in $C_i$ (or $C_{i,j}$)
to $U$, $V$, $W$, and $Z$\label{table1}}
\end{table}

Let $X\in \{U,V,W,Z\}$ and define $X_i=X\cap\lfloor C_i \rfloor_1$
and $X_{i,j}=X\cap\left\lfloor C_{i,j}\right\rfloor_1$ (for
example $V_{3,2}=V \cap\left\lfloor C_{3,2} \right\rfloor_1$). The
following relations can be easily obtained from (2) and Table 1:
$$\left\{\begin{array}{l}
|U_1|\leq |U|=q,\\
|V_2|+|V_{3,1}|+|V_{3,2}|+|V_{4,3}|\leq |V|=2q(q-1),\\
|W_{3,2}|+|W_{4,2}|+|W_{4,3}|\leq |W|=q(q-1),\\
|Z_{3,1}|+|Z_{4,1}|+|Z_{4,2}|+|Z_{4,3}|\leq |Z|=q(q-1)(q-2),\\
|C_1|=|U_1|,\,|C_2|=\frac{1}{2}|V_2|,\,|C_{3,1}|=\frac{1}{2}|V_{3,1}|= |Z_{3,1}|,\,|C_{3,2}|=\frac{1}{2}|V_{3,2}|=|W_{3,2}|,\\
|C_{4,1}|=\frac{1}{4}|Z_{4,1}|,\,|C_{4,2}|=\frac{1}{2}|W_{4,2}|=\frac{1}{2}|Z_{4,2}|,\,|C_{4,3}|=|V_{4,3}|=|W_{4,3}|=\frac{1}{2}|Z_{4,3}|.
\end{array}\right.$$

This information allows derivation of the first main result.
\begin{theorem}\label{bound1} Let $C$ be a code in $B_q^4$ with an even $q$ and $\rho(C)>2$.
Then
$$|C|\leq\frac{q^2(q+2)}{4}.$$
\end{theorem}
\begin{pf} It follows from the previous calculation that
$$\begin{array}{lll}
|C|&=&|C_1|+|C_2|+|C_{3,1}|+|C_{3,2}|+|C_{4,1}|+|C_{4,2}|+|C_{4,3}|\\
   &=&|U_1|+\frac{1}{2}|V_2|+\frac{1}{2}|V_{3,1}|+\frac{1}{2}|V_{3,2}|+\frac{1}{4}|Z_{4,1}|
   +\frac{1}{2}|Z_{4,2}|+\frac{1}{2}|Z_{4,3}|\\
   &=&|U_1|+\frac{1}{2}(|V_2|+|V_{3,1}|+|V_{3,2}|+|V_{4,3}|)-\frac{1}{2}|V_{4,3}|\\
   & &+\frac{1}{4}(|Z_{3,1}|+|Z_{4,1}|+|Z_{4,2}|+|Z_{4,3}|)-\frac{1}{4}|Z_{3,1}|+\frac{1}{4}|Z_{4,2}|+\frac{1}{4}|Z_{4,3}|\\
   &=&|U_1|+\frac{1}{2}(|V_2|+|V_{3,1}|+|V_{3,2}|+|V_{4,3}|)+\frac{1}{4}(|Z_{3,1}|+|Z_{4,1}|+|Z_{4,2}|+|Z_{4,3}|)\\
   & &+ \frac{1}{2}|C_{4,2}|- \frac{1}{4}|C_{3,1}|\\
   &\leq& |U|+\frac{1}{2}|V|+\frac{1}{4}|Z|+\frac{1}{2}|C_{4,2}|-
   \frac{1}{4}|C_{3,1}|.
\end{array}$$
Suppose that $(a,b_1,c_1,a)\in C_{4,2}$, $(a,b_2,c_2,a)\in
C_{4,2}$, and $(a,b_1,c_1,a)\neq (a,b_2,c_2,a)$. From the
condition $\rho(C)>2$, it follows that
$\{b_1,c_1\}\cap\{b_2,c_2\}=\emptyset$. Therefore
$$|C_{4,2}|\leq q\left\lfloor\frac{q-1}{2}\right\rfloor=\frac{q(q-2)}{2}.$$ Since $|C_{3,1}|\ge
0$, the result follows.
\end{pf}
\textbf{Remark 1.} Theorem \ref{bound1} improves the bound in
(\ref{bound0}) when $q$ is even and it gives the same bound when
$q$ is odd. But this bound is not sharp when $q$ is odd. Obtaining
a sharp bound when $q$ is odd seems to be a very difficult
problem.

\textbf{Remark 2.} The proof of Theorem \ref{bound1} can be used
in the construction of optimal codes in the following way. Let $C$
be a code in $B_q^4$ with $|C|=\frac{q^2(q+2)}{4}$. Then we should
have
$$\begin{array}{l}
|U_1|=|U|=q,\,|V_2|+|V_{3,1}|+|V_{3,2}|+|V_{4,3}|=|V|=2q(q-1),\,|C_{3,1}|=0,\\
|Z_{3,1}|+|Z_{4,1}|+|Z_{4,2}|+|Z_{4,3}|=|Z|=q(q-1)(q-2),\,|C_{4,2}|=\frac{q(q-2)}{2}.
\end{array}$$
Since $|C_{3,1}|=0$, we have $|V_{3,1}|=|Z_{3,1}|=0$. From the
fact that $|C_{4,2}|=\frac{q(q-2)}{2}$, we may assume that
$|C_{3,2}|=|C_{4,3}|=0$. Then $|V_{3,2}|=|V_{4,3}|=|Z_{4,3}|=0$.
In this case $|V_2|=|V|$. Hence
$$\begin{array}{l}
|C_1|=q,\,|C_2|=q(q-1),\,|C_3|=0,\\
|C_{4,1}|=\frac{q(q-1)(q-2)-q(q-2)}{4},
\,|C_{4,2}|=\frac{q(q-2)}{2}.
\end{array}$$
This means that we can obtain an optimal code if we include every
word of the types $(a, a, a, a)$ and $(a, a, b, b)$, the maximal
number of words of the type $(a, b, c, a)$, and generate words of
the type $(a, b, c, d)$ using all remaining words of length 3 of
the type $(a, b, c)$. Since $|U_1|=|U|$ and $|V_2|=|V|$ are always
possible, to construct an optimal code we only need to consider
combinations of elements in $W\cup Z$ that satisfy
$$|C_4|=\frac{q(q-1)(q-2)-q(q-2)}{4}+\frac{q(q-2)}{2}.$$

From these considerations, we can construct optimal codes in
$B_4^4$ (Table 2) and $B_6^4$ (Table 3) which coincide with the
construction in \cite{Leb}.

\begin{table}[h]
\centering
\begin{tabular}{|c|c|c|c|} \hline
0000 & 1111 & 2222 & 3333   \\
0011 & 0022 & 0033 & 1100   \\
1122 & 1133 & 2200 & 2211   \\
2233 & 3300 & 3311 & 3322   \\
0230 & 1231 & 2012 & 3013   \\
0321 & 2103 & 1302 & 3120   \\
\hline
\end{tabular}
\caption{An optimal code in $B_4^{4}$ \label{table2}} \vspace{0.2cm}
\centering
\end{table}

\begin{table}[h]
\centering
\begin{tabular}{|c|c|c|c|c|c|} \hline
0230 & 1231 & 2012 & 3013 & 4014 & 5015 \\
0450 & 1451 & 2452 & 3453 & 4234 & 5235 \\
0251 & 1304 & 2053 & 3105 & 4035 & 5102 \\
0342 & 1325 & 2140 & 3124 & 4120 & 5143 \\
0431 & 1503 & 2413 & 3520 & 4215 & 5321 \\
0524 & 1542 & 2504 & 3541 & 4302 & 5340 \\
\hline
\end{tabular}
\caption{The $C_4$ of an optimal code in $B_6^{4}$ \label{table3}}
\vspace{0.2cm} \centering
\end{table}

In the next two sections, we will construct an optimal code in
$B_q^{4}$ when $q$ is even. Let $x$ be a word of length 4 which
consists of pairwise distinct letters. The basic ingredients in
our construction are the codes $\langle x\rangle_{A_q^4}$
generated by $x$ in $A_q^4$ and $\langle x\rangle_{B_q^4}$
generated by $x$ in $B_q^4$. These codes will be defined below.
The code $\langle x\rangle_{A_q^4}$ was already used in \cite{L}.
The essence of our construction is to replace $\langle
x\rangle_{A_q^4}$ by $\langle x\rangle_{B_q^4}$ for some words
such that our construction satisfies the conditions of Remark 2.

Let $A_q^{n}$ be the set of all words in $B_q^{n}$ that have
pairwise distinct letters. For a word $x = (a_1,a_2,a_3,a_4)$ in
$A_q^{4}$, the codes $\langle x\rangle_{A_q^4}$ and $\langle
x\rangle_{B_q^4}$ are defined as follows:
$$\begin{array}{lll}
 \langle x\rangle_{A_q^4}&=&\{(a_1,a_2,a_3,a_4),
(a_1,a_4,a_3,a_2),
(a_2,a_4,a_1,a_3),\\
& & (a_3,a_4,a_1,a_2),(a_3,a_2,a_1,a_4),(a_4,a_2,a_3,a_1)\},\\
\langle x\rangle_{B_q^4}&=&\left(\langle
x\rangle_{A_q^4}\setminus\{(a_1,a_2,a_3,a_4),
(a_3,a_4,a_1,a_2)\}\right)\\
 & & \cup \{(a_1,a_3,a_4,a_1),
(a_2,a_3,a_4,a_2), (a_3,a_1,a_2,a_3), (a_4,a_1,a_2,a_4) \}.
\end{array}$$

The following lemma which describes the basic properties of these
codes under the deletion map can be easily verified.

\begin{lemma}\label{deletion1}
Let $x = (a_1,a_2,a_3,a_4)$ and $y = (b_1,b_2,b_3,b_4)$ be
distinct words of $A_q^4$, and $L(x)$ be the set of letters in
$x$. Then
\begin{enumerate}[(i)]
\item $\langle x\rangle_{A_q^4}$ and $\langle x\rangle_{B_q^4}$
are codes that are capable of correcting single deletions, \item
if $|L(x) \cap L(y)| \leq 2$, then $\langle x\rangle_{A_q^4} \cup
\langle y\rangle_{B_q^4}$ is a code that is capable of correcting
single deletions, \item if $\{a_1,a_2\} \cap \{b_1,b_2\} =
\emptyset$ or $\{a_3,a_4\} \cap \{b_3,b_4\} = \emptyset$, then
$\langle x\rangle_{B_q^4} \cup \langle y\rangle_{B_q^4}$ is a code
that is capable of correcting single deletions.
\end{enumerate}
\end{lemma}

\section{Construction of optimal codes in $B_q^4$ for $q\equiv 2 \hspace{0.2cm}\text{or}\ 4\,(\text{mod}\,6)$}

In this section, we will construct a code in $B_q^{4}$ for $q
\equiv 2\ \text{or}\ 4\,(\text{mod}\,6)$ that is capable of
correcting single deletions whose cardinality meets the upper
bound that was established in Theorem \ref{bound1}. Our
construction consists of two steps. In the first step, we
introduce the concept of the step property for a Steiner quadruple
system, and prove that there is a code in $B_q^{4}$ that is
capable of correcting single deletions whose cardinality meets the
upper bound in Theorem \ref{bound1}, under the assumption that
there is a Steiner quadruple system on $B_q$ that satisfies the
step property. In the next step, we follow the construction of
Hanani \cite{H} to show that there is a Steiner quadruple system
on $B_q$ that satisfies the step property for $q\equiv 2\
\text{or}\ 4\ (\text{mod}\ 6)$.

When $q=2$, we construct a code $C$ in $B_2^4$ with codewords
$$C=\left\{(0,0,0,0),\,(1,1,1,1),\,(0,0,1,1),\,(1,1,0,0)\right\},$$
by following the method in Remark 2. Since $|C|=4$, the code $C$
should be an optimal code.

Let $SQS(q)$ denote a Steiner quadruple system on $B_q$. Recall
that an $SQS(q)$ is a set of 4-element subsets of $B_q$, called
quadruples, with the property that every 3-element subset of $B_q$
is a subset of exactly one quadruple in the set. It is well known
\cite{LR} that there is a Steiner quadruple system on $B_q$ if and
only if $q\equiv 2\ \text{or}\ 4\ (\text{mod}\ 6)$. From now on,
we assume that $q\equiv 2\ \text{or}\ 4\ (\text{mod}\ 6)$ and that
$q\ge 4$.

To construct optimal codes, we define the step property.
\begin{definition}[The step property]
Let $SQS(q)$ be a Steiner quadruple system on $B_q$ and
$L_0<L_1<\cdots<L_{q-1}$ be a total order on
$B_q=\left\{L_0,L_1,\ldots,L_{q-1}\right\}$. Let $\{L_{2t},\,
_{2t+1},\,L_a,\,L_b\}$ be a quadruple in $SQS(q)$. We say that
$\{L_{2t},\,L_{2t+1},\,L_a,\,L_b\}$ satisfies the step property
with respect to the given order if either $a<2t,\,b< 2t$ or
$2t+1<a,\,2t+1<b$. We also say that $SQS(q)$ satisfies the step
property if every quadruple of the form
$\{L_{2t},\,L_{2t+1},\,L_a,\, L_b\}$ satisfies the step property
with respect to the given total order.
\end{definition}
If there is a Steiner quadruple system that has the step property,
optimal codes can be constructed using the following theorem.
\begin{theorem}\label{step1} Suppose that there is an $SQS(q)$
that satisfies the step property. Then
$$N(4,q,1)=\frac{q^2(q+2)}{4}.$$
\end{theorem}
\begin{pf}
Suppose that there is an $SQS(q)$ that satisfies the step
property. Without loss of generality, we may assume that
$0<1<\cdots<q-1$ is the total order which admits the step property
for $SQS(q)$. Let $\phi\,:\,SQS(q)\rightarrow B_q^4$ be the map
defined by $\phi\left(\{x,y,z,w\}\right)=(x,y,z,w)$ where
$x<y<z<w$, and
$$S(q)=\{\{2t,2t+1,a,b\}\in SQS(q)\ |\ 2t+1 <a,\,2t+1<b\}.$$ It
follows from the definition of a Steiner quadruple system and Lemma
\ref{deletion1} that the code
$$M=\left(\bigcup_{x\in\phi\left(SQS(q)\setminus S(q)\right)}\langle x\rangle_{A_q^4}\right)
\bigcup\left(\bigcup_{x\in\phi\left(S(q)\right)}\langle
x\rangle_{B_q^4}\right)$$ is capable of correcting single
deletions. $\left|S(q)\right|=\frac{q(q-2)}{8}$, so
$$|M|=6\cdot\left\{\frac{q(q-1)(q-2)}{24}-\frac{q(q-2)}{8}\right\}+8\cdot\frac{q(q-2)}{8}=\frac{q^2(q-2)}{4}.$$
We obtain a code $C$ from $M$ by adding all words of the types
$(a,a,a,a)$, and $(a,a,b,b)\ (a \neq b)$. It is easy to see that
$C$ is capable of correcting single deletions. Finally we have
$$|C|=q+q(q-1)+\frac{q^2(q-2)}{4}=\frac{q^2(q+2)}{4}.$$
\end{pf}

 Hanani \cite{H} inductively constructed an $SQS(q)$ for $q\equiv
2\ \text{or}\ 4\ (\text{mod}\ 6)$. Using Hanani's contruction, we
will prove that there exists an $SQS(q)$ with the step property
for all $q\equiv 2\ \text{or}\ 4\ (\text{mod}\ 6)$.

In subsequence sections, the following definitions and notations
will be used. For a natural number $n$, $B_n$ (resp. $\bar{B}_n$)
denotes the set $\{0,1,\ldots,n-1\}$ (resp. $\{1,2,\ldots,n\}$).
We introduce two partitions of unordered pairs of the set $B_{2m}$
\cite{H}.

We decompose the $m(2m-1)$ pairs $\{r,s\}$ from the set $B_{2m}$
into $2m-1$ systems $P_{\alpha}(m)\ (\alpha\in B_{2m-1})$, each
containing $m$ mutually disjoint pairs.

For $m\equiv 0\ (\text{mod}\ 2)$ we form the systems
$P_{\alpha}(m)$ as follows:
\begin{align*}
&P_{2\beta}(m)&=&\left\{\{2a,\,2a+2\beta+1\}\ |\
a\in B_m\right\}\ \left(\beta\in B_{\frac{m}{2}}\right),\\
&P_{2\beta+1}(m)&=&\left\{\{2a,\,2a-2\beta-1\}\ |\
a\in B_m\right\}\ \left(\beta\in B_{\frac{m}{2}}\right),\\
&P_{m+\gamma}(m)&=&\left\{\begin{array}{l} \{b,\,2\gamma-b\}\ |\
b\in B_{\gamma}\\
\{c,\,2m+2\gamma-c-2\}\ |\ 2\gamma+1\leq c\leq m+\gamma-2\\
\left\{2m-\frac{3}{2}-(-1)^{\gamma}\frac{1}{2},\gamma\right\}\\
\left\{
2m-\frac{3}{2}+(-1)^{\gamma}\frac{1}{2}m+\gamma-1\right\}\end{array}\right\}\
(\gamma\in B_{m-1}).
\end{align*}

For $m\equiv 1\ (\text{mod}\ 2)$ we let:
\begin{align*}
&P_{2\beta}(m)&=&\left\{\{2a,\,2a+2\beta+1\}\ |\
a\in B_m\right\}\ \left(\beta\in B_{\frac{m-1}{2}}\right),\\
&P_{2\beta+1}&=&\left\{\{2a,\,2a-2\beta-1\}\ |\
a\in B_m\right\}\ \left(\beta\in B_{\frac{m-1}{2}}\right),\\
&P_{m-1+\gamma}(m)&=&\left\{\begin{array}{l}\{b,\,2\gamma-b\}\ |\
b\in B_{\gamma}\\
\{c,2m+2\gamma-c\}\ |\ 2\gamma+1\leq c\leq m+\gamma-1\\
\{\gamma,m+\gamma\}
\end{array}\right\}\ (\gamma\in B_m).
\end{align*}

It can be easily verified that the pairs in every system are
mutually disjoint and no pair appears twice. Because the number of
pairs in the systems is $m(2m-1)$, it follows that every pair
appears in some system.

We shall also need another decomposition of the $m(2m-1)$ pairs from
$B_{2m}$ into $2m$ systems \textit{$\bar{P}_{\xi}(m)$} $(\xi\in
B_{2m})$ such that each of the $m$ systems $\bar{P}_\eta(m)$
$(\eta\in B_m)$ should contain $m-1$ mutually disjoint pairs not
containing the elements $2\eta$ and $2\eta+1$, and each of the other
$m$ systems should contain $m$ mutually disjoint pairs. We shall
form the system $\bar{P}_{\xi}(m)$ using the systems
$P_{\alpha}(m)$.

If $m\equiv 0\ (\text{mod}\ 2)$, it can easily be seen that
$$\begin{cases}
\{2\mu,\,4\mu+1\}\in P_{2\mu}(m)\ \left(\mu\in B_{\frac{m}{2}}\right),\\
\{2m-2-2\mu,\,2m-1-4\mu\}\in
P_{2\mu-1}(m)\ \left(\mu\in \bar{B}_{\frac{m-2}{2}}\right),\\
\{2m-2,\,0\}\in P_m(m),\\
\{2m-1,\,1\}\in P_{m+1}(m).
\end{cases}$$
Clearly, these pairs are mutually disjoint. We remove them from
their respective systems and use them to form a new system.

Performing the following permutation of the elements
$$\begin{pmatrix}
2\mu&4\mu+1&2m-2-2\mu&2m-1-4\mu&2m-2&0&2m-1&1\\
4\mu&4\mu+1&4\mu-2&4\mu-1&1&0&2m-1&2m-2\end{pmatrix}\ \left(\mu\in
\bar{B}_{\frac{m-2}{2}}\right),$$ we obtain the new systems
$\bar{P}_{\xi}(m)$ by a suitable reordering of the systems.

If $m\equiv 1\ (\text{mod}\ 2)$,
$$\begin{cases}
\{2\mu,\,4\mu+1\}\in P_{2\mu}(m)\ \left(\mu\in B_{\frac{m-1}{2}}\right),\\
\{2m-2-2\mu,\,2m-3-4\mu\}\in
P_{2\mu+1}(m)\ \left(\mu\in B_{\frac{m-1}{2}}\right),\\
\{m-1,\,2m-1\}\in P_{2m-2}(m).
\end{cases}$$
These pairs are again mutually disjoint.

By the permutation
$$\begin{pmatrix}
2\mu&4\mu+1&2m-2-2\mu&2m-3-4\mu&m-1&2m-1\\
2\mu&4\mu+1&4\mu+2&4\mu+3&2m-2&2m-1
\end{pmatrix}\ \left(\mu\in B_{\frac{m-1}{2}}\right)$$
of the elements and using the same procedure as in the case
$m\equiv 0\ (\text{mod} 2)$ we obtain the systems
$\bar{P}_{\xi}(m)$.

If a system $SQS(f)$ exists, we say that a quadruple system can be
formed from $B_f$ and write $f\in \mathcal{S}$. Similarly, if a
system $SQS(f)$ with the step property exists, we denote the
condition as $f \in \mathcal{SP}$.

If $f \in \mathcal{S}$, we shall use $\{x,y,z,t\}\in B_f$ to
denote any quadruple in $B_f$, that is, an element of $SQS(f)$. If
$f+1\in\mathcal{S}$, then we assume that $SQS(f+1)$ is formed by
the alphabet $B_f\cup \{A\}$ where $A$ is an additional element.
The quadruples that contain $A$ will be denoted by $\{A,u,v,w\}$.

The following was inductively shown by Hanani \cite{H}.
\begin{theorem}
If $q\equiv 2\ \text{or}\ 4\ (\text{mod}\ 6)$, then $q\in
\mathcal{S}$.
\end{theorem}
Using a similar argument as in \cite{H}, we will prove the
following.
\begin{theorem}\label{step2} Suppose that $q\ge 4$.
If $q\equiv 2\ \text{or}\ 4\ (\text{mod}\ 6)$, then $q\in
\mathcal{SP}$.
\end{theorem}
\begin{pf}
We will proceed by induction on $q$. Clearly, $SQS(4)$ has the
step property, hence $4 \in \mathcal{SP}$. As an induction step,
we will prove the following: Let $q\equiv 2\ \text{or}\ 4\
(\text{mod}\ 6)$. If $f \in \mathcal{SP}$ for every $f<q$
satisfying $f\equiv 2\ \text{or}\ 4\ (\text{mod}\ 6)$, then $q\in
\mathcal{SP}$. The proof will be given separately for each of the
following cases which evidently exhaust all the possibilities:
$$\begin{cases}
q\equiv 4\ \text{or}\ 8\ (\text{mod}\ 12),\\
q\equiv 4\ \text{or}\ 10\ (\text{mod}\ 18),\\
q\equiv 34\ (\text{mod}\ 36),\\
q\equiv 26\ (\text{mod}\ 36),\\
q\equiv 2\ \text{or}\ 10\ (\text{mod}\ 24)\ (q>2),\\
q\equiv 14\ \text{or}\ 38\ (\text{mod}\ 72).
\end{cases}$$

\textbf{Case I}\ : $q\equiv 4\ \text{or}\ 8\ (\text{mod}\ 12)$.\\
Let $q=2f$ where $f\equiv 2\ \text{or}\ 4\ (\text{mod}\ 6)$. Since
$f \in \mathcal{SP}$, there is an $SQS(f)$ with the step property.
Without loss of generality, we may assume that $0<1<\cdots<f-1$ is
the total order on $B_f$ which admits the step property for
$SQS(f)$. Let $N= \{(i,j)\ |\ i\in B_2,\,j \in B_f\}$ and
$\{x,y,z,t\}$ be any element in $SQS(f)$. The following quadruples
in $N$ form an $SQS(q)$ \cite{H} :
$$\left\{\begin{array}{lll}
Q_1&:&\{(a_1,x),(a_2,y),(a_3,z),(a_4,t)\}\ (a_1+a_2+a_3+a_4\equiv
0\
(\text{mod}\ 2)),\\
Q_2&:&\{(0,j),(0,j'),(1,j),(1,j')\}\ (j\neq j').
\end{array}\right.$$

We rename the letters of $N$ as follows: for $j\in B_f$,
\begin{equation}\label{rename1}(0,j)\rightarrow j,\,
(1,j)\rightarrow f+j.\end{equation} Note that the solutions of
$a_1+a_2+a_3+a_4\equiv 0\ (\text{mod}\ 2)$ are
$$
(a_1,a_2,a_3,a_4)=
\begin{cases}(0,0,0,0),\,(0,0,1,1),\,(0,1,0,1),\,(0,1,1,0),\\
             (1,0,0,1),\,(1,0,1,0),\,(1,1,0,0),\,(1,1,1,1).\end{cases}$$
To investigate the step property on $SQS(q)$, we only need to
consider the following quadruples in each type: for $t\in
B_{\frac{t}{2}}$,
$$\left\{\begin{array}{lll}
Q_1 &:& \{(0,2t),\,(0,2t+1),\,(0,x),\,(0,y)\},\,\{(0,2t),\,(0,2t+1),\,(1,x),\,(1,y)\}, \\
\mbox{ } &&
\{(1,2t),\,(1,2t+1),\,(0,x),\,(0,y)\},\,\{(1,2t),\,(1,2t+1),\,(1,x),\,(1,y)\},\\
Q_2 &:& \{(0,2t),\,(0,2t+1),\,(1,2t),\,(1,2t+1)\}.
\end{array}\right.$$
Note that $\{2t,\,2t+1,\,x,\,y\}$ is a quadruple in $SQS(f)$. Since
$SQS(f)$ satisfies the step property with respect to the order
$0<1<\cdots<f-1$, it follows from (\ref{rename1}) that $SQS(q)$
satisfies the step property. Therefore $q\in \mathcal{SP}$.

\textbf{Case II}\ : $q\equiv 4\ \text{or}\ 10\ (\text{mod}\
18)$.\\
Let $q=3f+1$ where $f\equiv 1\ \text{or}\ 3\ (\text{mod}\ 6)$.
Since $f+1\equiv 2\ \text{or}\ 4\ (\text{mod}\ 6)$, we obtain
$f+1\in \mathcal{SP}$ by induction. Hence we may assume that
$SQS(f+1)$ satisfies the step property with respect to the order
$0<1<\cdots<f-1<A$. In accordance with the notation in \cite{H},
we define $N=\{(i,j),\,A\ |\ i\in B_3,\,j \in B_f\}$. Also, let
$\{x,y,z,t\}$ and $\{A,u,v,w\}$ be quadruples in $B_f$ and $B_f
\cup \{A\}$, respectively. Note that both of them are elements of
$SQS(f+1)$. According to \cite{H}, the following quadruples in $N$
form an $SQS(q)$:
$$\left\{\begin{array}{lll}
Q_1&:&\{(a_1,x),(a_2,y),(a_3,z),(a_4,t)\}\ (a_1+a_2+a_3+a_4\equiv
0\
(\text{mod}\ 3)),\\
Q_2&:&\{A,(b_1,u),(b_2,v),(b_3,w)\}\ (b_1+b_2+b_3\equiv 0\
(\text{mod}\
3)),\\
Q_3&:&\{(i,u),(i,v),(i+1,w),(i+2,w)\},\\
Q_4&:&\{(i,j),(i,j'),(i+1,j),(i+1,j')\}\ (j\neq j'),\\
Q_5&:&\{A,(0,j),(1,j),(2,j)\}.
\end{array}\right.$$
In cases $Q_3$ and $Q_4$, calculations are performed modulo 3.

We then rename letters in $N$ as follows: for $t\in
B_{\frac{f-1}{2}}$,
\begin{equation}\label{rename2}\left\{\begin{array}{ll}
(0,2t)\rightarrow 6t,&(0,2t+1)\rightarrow 6t+1,\\
(1,2t)\rightarrow 6t+2,&(1,2t+1)\rightarrow 6t+3,\\
(2,2t)\rightarrow 6t+4,&(2,2t+1)\rightarrow 6t+5\\
(0,f-1)\rightarrow 3f-3,&(1,f-1)\rightarrow
3f-2,\\
(2,f-1)\rightarrow 3f-1,&A\rightarrow 3f.
\end{array}\right.\end{equation}
To investigate the step property, we only need to consider the
following quadruples of each type: for $t\in B_{\frac{f-1}{2}}$
and $i\in B_4$,
$$\left\{\begin{array}{lll}
Q_1&:&\{(0,2t),(0,2t+1),(0,x),(0,y)\}, \{(0,2t),(0,2t+1),(1,x),(2,y)\},\\
\mbox{ } && \{(1,2t),(1,2t+1),(0,x),(1,y)\}, \{(1,2t),(1,2t+1),(2,x),(2,y)\},\\
\mbox{ } && \{(2,2t),(2,2t+1),(0,x),(2,y)\},
\{(2,2t),(2,2t+1),(1,x),(1,y)\},\\
Q_2&:&\{(1,2t),(1,2t+1),(1,x),A\},\,\{(0,x),(1,y),(2,f-1),A\},\\
Q_3&:&\{(0,2t),(0,2t+1),(1,x),(2,x)\},\{(1,2t),(1,2t+1),(2,x),(0,x)\},\\
\mbox{}
&&\{(2,2t),(2,2t+1),(0,x),(1,x)\},\{(2,x),(2,y),(0,f-1),(1,f-1)\},\\
Q_4&:&\{(i,2t),(i,2t+1),(i+1,2t),(i+1,2t+1)\},\\
Q_5&:&\{A,(0,f-1),(1,f-1),(2,f-1)\}.
\end{array}\right.$$
Note that $\{2t,2t+1,a,b\}$, $\{2t,2t+1,a,A\}$, and
$\{a,b,f-1,A\}$ are quadruples in $SQS(f+1)$. Since $SQS(f+1)$ on
$B_f\cup\{A\}$ satisfies the step property with respect to the
order $0<1<\cdots<f-1<A$, it follows from (\ref{rename2}) that
$SQS(q)$ satisfies the step property. Therefore $q\in
\mathcal{SP}$.

\textbf{Case III}\ : $q\equiv 34\ (\text{mod}\ 36)$.\\
Let $q=3f+4$ where $f \equiv 10\ (\text{mod}\ 12)$, and denote
$f=12k+10$. By induction, we may assume that $f+4\in
\mathcal{SP}$. In accordance with the notation in \cite{H}, we can
assume that there is an $SQS(f+4)$ on the alphabet
$B_f\cup\left\{(A,0),\,(A,1),\,(A,2),\,(A,3)\right\}$. Defining
$L_t=2t,\,L_{\frac{f}{2}+t}=2t+1\ (t\in B_{\frac{f}{2}})$, we also
assume that given $SQS(f+4)$ satisfies the step property with
respect to the order
$$L_0<L_1<\cdots<L_{f-1}<(A,0)<(A,1)<(A,2)<(A,3).$$

Let $N=\left\{(i,j), (A,h) \bigm | i\in B_3,\,j \in B_f,\,h\in
B_4\right\}$ and $\{x,y,z,t\}$ be any quadruple in $SQS(f+4)$. The
following quadruples in $N$ form an $SQS(q)$ \cite{H}:
$$\left\{\begin{array}{lll}
Q_1&:&\{(A,0),(A,1),(A,2),(A,3)\},\\
Q_2&:&\{(i,x),(i,y),(i,z),(i,t)\}\ (\text{the quadruple in}\ Q_1\ \text{excluded}),\\
Q_3&:&\{(A,a_1),(0,a_2),(1,a_3),(2,a_4)\} (a_1+a_2+a_3+a_4\equiv
0\
(\text{mod}\ f)),\\
Q_4&:&\{(i+2,b_3),(i,b_1+2k+1+i(4k+2)-d),\\
    & & (i,b_1+2k+2+i(4k+2)+d),(i+1,b_2)\}\\
   & &\text{where}\ b_1+b_2+b_3\equiv 0\ (\text{mod}\ f)\ \text{and}\ d\in B_{2k+1},\\
Q_5&:&\{(i,r_{\alpha}),(i,s_{\alpha}),(i+1,r'_{\alpha}),(i+1,s'_{\alpha})\}\
\text{where}\ \{r_{\alpha},s_{\alpha}\}\ \text{and}\
   \{r'_{\alpha},s'_{\alpha}\} \\
   & &\text{are (equal or different)
   pairs in}\ P_{\alpha}(6k+5)\ (4k+2\leq \alpha\leq 12k+8).
\end{array}\right.$$
In case $Q_2$, we define $(i,(A,h))=(A,h)$ for all $i\in B_3$ and
$h\in B_3$. For both cases $Q_4$ and $Q_5$, calculations are
conducted modulo 3 and $f$ for the first and second coordinates,
respectively.

We rename letters in $N$ as follows:
\begin{equation}\label{rename3}\left\{\begin{array}{l}
(i,L_t)\rightarrow if+t\ (i\in B_3,\,t\in B_f),\\
(A,h)\rightarrow 3f+h\ (h\in B_4).
\end{array}\right.
\end{equation}
To investigate the step property, we first consider the following
quadruples of each type except $Q_5$: for $t\in B_{\frac{f}{2}}$,
$i\in B_3$, and $h\in B_2$,
$$\left\{\begin{array}{lll}
Q_1&:&\{(A,0),(A,1),(A,2),(A,3)\},\\
Q_2&:&\{(i,L_{2t}),(i,L_{2t+1}),(i,x),(i,y)\},\{(i,x'),(i,y'),(A,2h),(A,2h+1)\},\\
Q_3,\,Q_4&:& \text{no quadruples to consider}.
\end{array}\right.$$
Note that $\{(A,0),(A,1),(A,2),(A,3)\}$, $\{L_{2t}, L_{2t+1},x,y\}$,
and $\{x',y',(A,2h),(A,2h+1)\}$ are quadruples in $SQS(f+4)$. Since
$SQS(f+4)$ satisfies the step property with respect to the given
order, it can be easily checked by (\ref{rename3}) that the
quadruples above satisfy the step property.

In $Q_5$, the following quadruples should be considered: for $i\in
B_3$,
$$\{(i,L_{2t}),(i,L_{2t+1}),(i+1,x),(i+1,y)\},\,\{(i,x'),(i,y'),(i+1,L_{2t}),(i+1,L_{2t+1})\},$$
where $\{2t,2t+1\}$ and $\{x,y\}$ are pairs in $P_{\alpha}(6k+5)\
(4k+2\leq\alpha\leq 12k+8)$. These quadruples also satisfy the step
property. Therefore $q\in \mathcal{SP}$.

\textbf{Case IV}\ : $q\equiv 26\ (\text{mod}\ 36)$.\\
This case is similar to Case III. Let $q=3f+2$ where $f \equiv 8\
(\text{mod}\ 12)$, and denote $f=12k+8$. By induction, we have
$f+2\in \mathcal{SP}$. We may assume that there is an $SQS(f+2)$
on the alphabet $B_f\cup\{(A,0),(A,1)\}$. Defining
$L_t=2t,\,L_{\frac{f}{2}+t}=2t+1\ \left(t\in
B_{\frac{f}{2}}\right)$, we also assume that the given $SQS(f+2)$
satisfies the step property with respect to the order
$$L_0<L_1<\cdots<L_{f-1}<(A,0)<(A,1).$$

Let $N=\{(i,j),\,(A,h)\ |\ i\in B_3,\,j \in B_f,\,h\in B_2\}$ and
$\{x,y,z,t\}$ be any quadruple in $SQS(f+2)$. The following
quadruples in $N$ form an $SQS(q)$ \cite{H}:
$$\left\{\begin{array}{lll}
Q_1&:&\{(i,x),(i,y),(i,z),(i,t)\},\\
Q_2&:&\{(A,a_1),(0,a_2),(1,a_3),(2,a_4)\}\ (a_1+a_2+a_3+a_4\equiv
0\
(\text{mod}\ f)),\\
Q_3&:&\{(i+2,b_3),(i,b_1+2k+1+i(4k+2)-d),\\
   & &(i,b_1+2k+2+i(4k+2)+d),(i+1,b_2)\}\\
   & &\text{where}\ b_1+b_2+b_3\equiv 0\ (\text{mod}\ f)\ \text{and}\ d\in B_{2k+1},\\
Q_4&:&\{(i,r_{\alpha}),(i,s_{\alpha}),(i+1,r'_{\alpha}),(i+1,s'_{\alpha})\}\
\text{where}\ \{r_{\alpha},s_{\alpha}\}\ \text{and}\
   \{r'_{\alpha},s'_{\alpha}\}\\
   & &\ \text{are (equal or different) pairs in}\
   P_{\alpha}(6k+4)\ (4k+2\leq\alpha\leq 12k+6).
\end{array}\right.$$
In case $Q_1$, we define $(i,(A,h))=(A,h)$ for all $i\in B_3$ and
$h\in B_4$. For both cases $Q_3$ and $Q_4$, calculations are
performed modulo 3 and $f$ for the first and second coordinates,
respectively.

We again rename letters in $N$ as follows:
\begin{equation}\label{rename4}
\left\{\begin{array}{l} (i,L_t)\rightarrow
if+t\ (i\in B_3,\,t\in B_f),\\
(A,h)\rightarrow 3f+h\ (h\in B_2).
\end{array}\right.
\end{equation}
To investigate the step property, we will first consider the
following quadruples of each type except $Q_4$ : for $t\in
B_{\frac{f}{2}}$ and $i\in B_3$,
$$\left\{\begin{array}{lll}
Q_1&:&\{(i,L_{2t}),(i,L_{2t+1}),(i,a),(i,b)\},\{(i,a'),(i,b'),(A,0),(A,1)\},\\
Q_2,\,Q_3&:& \text{no quadruples to consider}.
\end{array}\right.$$
Note that $\{L_{2t},L_{2t+1},a,b\}$ and $\{a',b',(A,0),(A,1)\}$ are
quadruples in $SQS(f+2)$. Since $SQS(f+2)$ satisfies the step
property with respect to the order above, it can be easily checked
by (\ref{rename4}) that the quadruples above satisfy the step
property.

In case $Q_4$, the following quadruples should be considered: for
$i\in B_3$,
$$\{(i,L_{2t}),(i,L_{2t+1}),(i+1,a),(i+1,b)\},\,\{(i,a),(i,b),(i+1,L_{2t}),(i+1,L_{2t+1})\},$$
where $\{L_{2t},L_{2t+1}\}$ and $\{a,b\}$ are pairs in
$P_{\alpha}(6k+4)\ (4k+2\leq\alpha\leq 12k+6)$. These quadruples
also satisfy the step property. Therefore $q\in\mathcal{SP}$.

\textbf{Case V}\ : $q\equiv 2\ \text{or}\ 10\ (\text{mod}\ 24)$
$(q>2)$.\\
Let $q=4f+2$ where $f\equiv 0\ \text{or}\ 2\ (\text{mod}\ 6)$
$(f>0)$, and denote $f=2k$. By induction, we have $f+2\in
\mathcal{SP}$. Hence we may assume that $SQS(f+2)$ on
$B_f\cup\{(A,0),(A,1)\}$ satisfies the step property  with respect
to the order
$$0<1<\cdots<f-1<(A,0)<(A,1).$$ Let $N=\{(h,i,j),\,(A,l)\ |\ h\in
B_2,\,i\in B_2,\,j \in B_f,\,l\in B_2\}$ and $\{x,y,z,t\}$ be any
quadruple in $SQS(f+2)$. The following quadruples in $N$ form an
$SQS(q)$ \cite{H}: assuming that $c_1+c_2+c_3\equiv 0\ (\text{mod}\
k)$ and $\epsilon\in B_2$,
$$\left\{\begin{array}{lll}
Q_1&:&\{(h,i,x),(h,i,y),(h,i,z),(h,i,t)\},\\
Q_2&:&\{(A,l),(0,0,2c_1),(0,1,2c_2-\epsilon),(1,\epsilon,2c_3+l)\},\\
Q_3&:&\{(A,l),(0,0,2c_1+1),(0,1,2c_2-1-\epsilon),(1,\epsilon,2c_3+1-l)\},\\
Q_4&:&\{(A,l),(1,0,2c_1),(1,1,2c_2-\epsilon),(0,\epsilon,2c_3+1-l)\},\\
Q_5&:&\{(A,l),(1,0,2c_1+1),(1,1,2c_2-1-\epsilon),(0,\epsilon,2c_3+l)\},\\
Q_6&:&\{(h,0,2c_1+\epsilon),(h,1,2c_2-\epsilon),(h+1,0,\bar{r}_{c_3}),(h+1,0,\bar{s}_{c_3})\}\\
   & &\text{where}\ \{\bar{r}_{c_3},\bar{s}_{c_3}\}\ \text{are pairs in}\
   \bar{P}_{c_3}(k),\\
Q_7&:&\{(h,0,2c_1-1+\epsilon),(h,1,2c_2-\epsilon),(h+1,1,\bar{r}_{c_3}),(h+1,1,\bar{s}_{c_3})\},\\
Q_8&:&\{(h,0,2c_1+\epsilon),(h,1,2c_2-\epsilon),(h+1,1,\bar{r}_{k+c_3}),(h+1,1,\bar{s}_{k+c_3})\},\\
Q_9&:&\{(h,0,2c_1-1+\epsilon),(h,1,2c_2-\epsilon),(h+1,0,\bar{r}_{k+c_3}),(h+1,0,\bar{s}_{k+c_3})\},\\
Q_{10}&:&\{(h,0,r_{\alpha}),(h,0,s_{\alpha}),(h,1,r'_{\alpha'}),(h,1,s'_{\alpha})\}\
\text{where}\ \{r_{\alpha},s_{\alpha}\}\ \text{and}\
    \{r'_{\alpha'},s'_{\alpha}\}
      \\
    & & \text{are (equal or different) pairs in}\ P_{\alpha}(k)\ (\alpha\in
    B_{f-1}).
\end{array}\right.$$
Note that if a quadruple in $Q_1$ contains an element of the form
$(h,i,(A,l))$, we simply denote it by $(A,l)$. For each case,
calculations are conducted modulo 2 and $f$ for the first and
third coordinates, respectively.

We rename letters in $N$ as follows: for $j\in B_f$,
\begin{equation}\label{rename5}\left\{\begin{array}{lll}
(0,0,j)\rightarrow j,&(0,1,j)\rightarrow f+j,\\
(1,0,j)\rightarrow 2f+j,&(1,1,j)\rightarrow 3f+j,\\
(A,0)\rightarrow 4f,&(A,1)\rightarrow 4f+1.
\end{array}\right.\end{equation}
To investigate the step property, the following quadruples of each
type should be considered, except $Q_6 - Q_{10}$ : for $t\in
B_{\frac{f}{2}}$, $h\in B_2$, and $i\in B_2$,
$$\left\{\begin{array}{lll}
Q_1&:&\{(h,i,2t),(h,i,2t+1),(h,i,a),(h,i,b)\},\,\{(A,0),(A,1),(h,i,a),(h,i,b)\},\\
Q_2-Q_5&:& \text{no quadruples to consider}.
\end{array}\right.$$
In case $Q_1$, the quadruples $\{2t,2t+1,a,b\}$ and
$\{(A,0),(A,1),a,b\}$ are in $SQS(f+2)$. Since $SQS(f+2)$
satisfies the step property with respect to the given order, it
can be easily checked by (\ref{rename5}) that the quadruples above
satisfy the step property.

In $Q_6$, the following quadruples are to be considered: for $h\in
B_2$ and $\epsilon\in B_2$,
$$\{(h,0,2c_1+\epsilon),(h,1,2c_2-\epsilon),(h+1,0,2t),(h+1,0,2t+1)\}.$$
These quadruples also satisfy the step property. Cases
$Q_7-Q_{10}$ are similar to $Q_6$. As a result,
$q\in\mathcal{SP}$.

\textbf{Case VI}\ : $q\equiv 14\ \text{or}\ 38\ (\text{mod}\ 72)$.\\
Let $q=12f+2$ where $f\equiv 1\ \text{or}\ 3\ (\text{mod}\ 6)$. As
the first step, we will prove that $14\in\mathcal{S}$ and $38
\in\mathcal{S}$. According to \cite{denny}, we can construct an
$SQS(14)$ (Table 4).
\begin{table}[h]
\centering
\begin{tabular}{|c|c|c|c|c|c|c|}\hline
\{0,1,2,4\}&\{13,9,12,11\}&\{12,1,4,3\}&\{12,8,5,1\}&\{5,6,10,11\}&\{3,5,13,8\}&\{4,5,7,8\}\\
\{7,10,13,12\}&\{0,8,9,11\}&\{13,2,5,4\}&\{13,9,6,2\}&\{6,0,11,12\}&\{4,6,7,9\}&\{9,12,4,0\}\\
\{1,2,3,5\}&\{7,3,6,5\}&\{3,4,10,11\}&\{7,10,0,3\}&\{0,1,12,13\}&\{5,0,8,10\}&\{10,13,5,1\}\\
\{2,3,4,6\}&\{1,9,10,12\}&\{9,12,2,5\}&\{8,11,1,4\}&\{1,2,13,7\}&\{6,1,9,11\}&\{11,7,6,2\}\\
\{3,4,5,0\}&\{2,10,11,13\}&\{6,1,13,8\}&\{0,2,7,9\}&\{2,3,7,8\}&\{0,2,10,12\}&\{12,8,0,3\}\\
\{4,5,6,1\}&\{3,11,12,7\}&\{4,5,11,12\}&\{1,3,8,10\}&\{3,4,8,9\}&\{5,6,8,9\}&\{13,9,1,4\}\\
\{5,6,0,2\}&\{4,12,13,8\}&\{5,6,12,13\}&\{2,4,9,11\}&\{13,9,0,3\}&\{8,11,3,6\}&\{7,10,2,5\}\\
\{6,0,1,3\}&\{5,13,7,9\}&\{6,0,13,7\}&\{3,5,10,12\}&\{7,10,1,4\}&\{3,5,9,11\}&\{4,6,10,12\}\\
\{8,11,7,13\}&\{6,7,8,10\}&\{0,1,7,8\}&\{4,6,11,13\}&\{8,11,2,5\}&\{6,0,9,10\}&\{5,0,11,13\}\\
\{9,12,8,7\}&\{8,4,0,6\}&\{1,2,8,9\}&\{5,0,12,7\}&\{9,12,3,6\}&\{0,1,10,11\}&\{6,1,12,7\}\\
\{10,13,9,8\}&\{9,5,1,0\}&\{2,3,9,10\}&\{4,5,9,10\}&\{10,13,4,0\}&\{1,2,11,12\}&\{0,2,13,8\}\\
\{11,7,10,9\}&\{10,6,2,1\}&\{10,13,3,6\}&\{12,8,6,2\}&\{11,7,5,1\}&\{2,3,12,13\}&\{1,3,7,9\}\\
\{12,8,11,10\}&\{11,0,3,2\}&\{11,7,4,0\}&\{1,3,11,13\}&\{2,4,12,7\}&\{3,4,13,7\}&\{2,4,8,10\}\\\hline
\end{tabular}
\caption{An SQS(14)\label{table4}}\centering
\end{table}

Defining $N'=\{(i,j),\,(A,h)\ |\ i\in B_3,\,j \in B_{12},\,h\in
B_2\}$ to be a set of 38 elements, we will show that
$38\in\mathcal{S}$. Let $\{x',y',z',t'\}$ be any quadruple in
$SQS(14)$. The following quadruples in $N'$ form an $SQS(38)$
\cite{H}: assuming that $b_1+b_2+b_3\equiv 0\ (\text{mod}\ 12)$,
$\epsilon\in B_2$, $g\in B_6$, and $e\in B_4$,
$$\left\{\begin{array}{lll}
Q_1&:&\{(i,x'),(i,y'),(i,z'),(i,t')\},\\
Q_2&:&\{(A,h),(0,b_1),(1,b_2),(2,b_3+3h)\},\\
  Q_3&:&\{(i,b_1+4+i),(i,b_1+7+i),(i+1,b_2),(i+2,b_3)\},\\
Q_4&:&\{(i,j),(i+1,j+6\epsilon),(i+2,6\epsilon-2j+1),(i+2,6\epsilon-2j-1)\},\\
Q_5&:&\{(i,j),(i+1,j+6\epsilon),(i+2,6\epsilon-2j+2),(i+2,6\epsilon-2j-2)\},\\
Q_6&:&\{(i,j),(i+1,j+6\epsilon-3),(i+2,6\epsilon-2j+1),(i+2,6\epsilon-2j+2)\},\\
Q_7&:&\{(i,j),(i+1,j+6\epsilon+3),(i+2,6\epsilon-2j-1),(i+2,6\epsilon-2j-2)\},\\
Q_8&:&\{(i,j),(i,j+6),(i+1,j+3\epsilon),(i+1,j+6+3\epsilon)\},\\
Q_9&:&\{(i,2g+3\epsilon),(i,2g+6+3\epsilon),(i',2g+1),(i',2g+5)\}\ (i'\neq i),\\
Q_{10}&:&\{(i,2g+3\epsilon),(i,2g+6+3\epsilon),(i',2g+2),(i',2g+4)\},\\
Q_{11}&:&\{(i,j),(i,j+1),(i+1,j+3e),(i+1,j+3e+1)\},\\
Q_{12}&:&\{(i,j),(i,j+2),(i+1,j+3e),(i+1,j+3e+2)\},\\
Q_{13}&:&\{(i,j),(i,j+4),(i+1,j+3e),(i+1,j+3e+4)\},\\
Q_{14}&:&\{(i,r_{\alpha}),(i,s_{\alpha}),(i',r'_{\alpha}),(i',s'_{\alpha})\}\
\text{where}\ \{r_{\alpha},s_{\alpha}\}\ \text{and}\
      \{r'_{\alpha},s'_{\alpha}\}\ \\
      & &\text{are (equal or different) pairs in}\
      P_{\alpha}(6)\ (4\leq\alpha\leq 5).
\end{array}\right.$$
In case $Q_1$, we define $(i,(A,h))=(A,h)$ for all $i\in B_3$ and
$h\in B_2$. For each case, calculations are conducted modulo 3 and
12 for the first and second coordinates, respectively.

Now we construct $SQS(q)$ for $q\equiv 14\ \text{or}\ 38\
(\text{mod}\ 72)$, that is, $f\equiv 1\ \text{or}\ 3\ (\text{mod}\
6)$. Since $f\equiv 1\ \text{or}\ 3\ (\text{mod}\ 6)$, we can
assume that $f+1\in \mathcal{S}$. In this case, the alphabet of
$SQS(f+1)$ is $B_f\cup\{A\}$ as we assumed. Let
$N=\{(i,j),\,(A,h)\ |\ i\in B_f,\,j\in B_{12},\,h\in B_2\}$ be a
set of $12f+2$ elements and $\{A,u,v,w\}$ be any quadruple in
$SQS(f+1)$ that contains $A$. The following quadruples from $N$
form an $SQS(q)$ \cite{H}:
$$\left\{\begin{array}{lll}
R_1&:&\{(i,x'),(i,y'),(i,z'),(i,t')\},\\
R_2&:&\begin{cases}\{(A,h),(u,b_1),(v,b_2),(w,b_3+3h)\}\
(b_1+b_2+b_3\equiv
0\ (\text{mod}\ 12)),\\
\{(u,\alpha_1),(v,\alpha_2),(w,\alpha_3),(w,\alpha_4)\},\\
\{(i,\beta_1),(i,\beta_2),(i',\beta_3),(i',\beta_4)\}\
\left(\{i',\,i\}\subset\{u,\,v,\,w\}\ \text{and}\ i'\neq
i\right).\end{cases}
\end{array}\right.$$
In case $R_2$, $\alpha_{\nu}$ and $\beta_{\nu}$
$(\nu\in\bar{B}_4)$ are to be replaced by the second indices of
$Q_3-Q_{14}$, corresponding to the first indices 0, 1, 2 for $u$,
$v$, $w$, respectively. Note that $i$ and $i'$ define uniquely a
$\{u,v,w\}$ in which they are contained. Therefore they may be
considered as two indices from $\{u,v,w\}$.
$$\begin{array}{lll}
R_3&:&\{(x,a_1),(y,a_2),(z,a_3),(t,a_4)\}\ (a_1+a_2+a_3+a_4\equiv
0\ (\text{mod}\ 12)).
\end{array}$$

From now on, we will show that $q\in \mathcal{SP}$ for $q\equiv
14\ \text{or}\ 38\ (\text{mod}\ 72)$. In the case that $q = 14$,
we rename the letters of $B_{14}$ in Table 4 as follows:
\begin{equation*}\left\{\begin{array}{ll}
t\rightarrow L_{2t}\ (t\in B_6),\\
t\rightarrow L_{2t-13}\ (7\leq t\leq 12),\\
6+7i\rightarrow L_{i+12}=(A,i)\ (i\in B_2)
\end{array}\right.\end{equation*}
Assuming $L_0<L_1<\cdots<L_{13}$, the following quadruples satisfy
the step property:
\begin{center}
\begin{tabular}[h]{lllllll}
&\{0,7,1,8\},&\{0,7,2,9\},&\{0,7,3,10\},&\{0,7,4,11\},&\{0,7,5,12\},&\{0,7,6,13\},\\
\{1,8,0,7\},&&\{1,8,2,9\},&\{1,8,3,10\},&\{1,8,4,11\},&\{1,8,5,12\},&\{1,8,6,13\},\\
\{2,9,0,7\},&\{2,9,1,8\},&&\{2,9,3,10\},&\{2,9,4,11\},&\{2,9,5,12\},&\{2,9,6,13\},\\
\{3,10,0,7\},&\{3,10,1,8\},&\{3,10,2,9\},&&\{3,10,4,11\},&\{3,10,5,12\},&\{3,10,6,13\},\\
\{4,11,0,7\},&\{4,11,1,8\},&\{4,11,2,9\},&\{4,11,3,10\},&&\{4,11,5,12\},&\{4,11,6,13\},\\
\{5,12,0,7\},&\{5,12,1,8\},&\{5,12,2,9\},&\{5,12,3,10\},&\{5,12,4,11\},&&\{5,12,6,13\},\\
\{6,13,0,7\},&\{6,13,1,8\},&\{6,13,2,9\},&\{6,13,3,10\},&\{6,13,4,11\},&\{6,13,5,12\}.&
\end{tabular}
\end{center}
This shows that $14\in\mathcal{SP}$.

In the case that $q=38$, we rename letters of $N'$ as follows:
\begin{equation}\label{rename6}\left\{\begin{array}{l}
(i,L_t)\rightarrow 12i+t\ (i\in B_3,\,t\in B_{12}),\\
(A,h)\rightarrow 36+h\ (h\in B_2).
\end{array}\right.\end{equation}
To investigate the step property of $SQS(38)$, we will consider
quadruples of each type.

In case $Q_1$, the following quadruples should be considered: for
$i\in B_3$,
$$\{(i,L_{2t}),(i,L_{2t+1}),(i,a),(i,b)\},\,\{(i,a),(i,b),(A,0),(A,1)\}.$$
Since $\{L_{2t},L_{2t+1},a,b\}$ and $\{a,b,(A,0),(A,1)\}$ are in
$SQS(14)$ and satisfy the step property, these quadruples also
satisfy the step property by (\ref{rename6}).

In cases $Q_2-Q_{13}$, there are no quadruples to consider.

In case $Q_{14}$, the following quadruples should be considered:
$$\{(i,L_{2t}),(i,L_{2t+1}),(i',a),(i',b)\}.$$
By (\ref{rename6}), these quadruples satisfy the step property.
Therefore $38\in \mathcal{SP}$.

Finally, consider general cases. First, rename letters in $N$ as
follows:
\begin{equation}\label{rename61}\left\{\begin{array}{l}
(i,L_t)\rightarrow 12i+t\ (i\in B_f,\,t\in B_{12}),\\
(A,h)\rightarrow 12f+h\ (h\in B_2).
\end{array}\right.\end{equation}
From now on, we will investigate the step property of quadruples
of each type.
\paragraph{$R_1$)}
$\{(i,L_{2t}),(i,L_{2t+1}),(i,a),(i,b)\},\,\{(i,a),(i,b),(A,0),(A,1)\}.$\\
Since $\{L_{2t},L_{2t+1},a,b\}$ and $\{a,b,(A,0),(A,1)\}$ satisfy
the step property, these quadruples also satisfy the step property
by (\ref{rename61}).
\paragraph{$R_2$)}
$\{(i,L_{2t}),(i,L_{2t+1}),(i',a),(i',b)\}.$\\ By the step
property of the constructed $SQS(38)$, these quadruples satisfy
the step property by (\ref{rename61}).
\paragraph{$R_3$)} In this case, there are no quadruples to
consider.

Therefore $q\in\mathcal{SP}$
\end{pf}

We constructed an optimal code in $B_2^4$, so if we use Theorems
\ref{step1} and \ref{step2}, we can deduce the following theorem.
\begin{theorem}\label{Steiner}
If  $q \equiv 2\ \text{or}\ 4\,(\text{mod}\ 6)$, then
$N(4,q,1)=\frac{q^2(q+2)}{4}$.
\end{theorem}

\textbf{Remark 3.} The authors of \cite{ZZ} informed us that every
Steiner quadruple system in \cite{ZZ} satisfies the step property.
It is not known whether every Steiner quadruple system satisfies
the step property.

\section{Construction of optimal codes in $B_q^4$ for $q\equiv 0\,(\text{mod}\,6)$}

In this section, we will construct a code in $B_q^{4}$ for $q
\equiv 0\, (\text{mod}\,6)$ that is capable of correcting single
deletions whose cardinality meets the upper bound that was
established in Theorem \ref{bound1}. Since there dose not exist a
Steiner quadruple system for any alphabet of this size, we will
use a group divisible system. We divide our construction into two
steps. In the first step, we prove that an optimal code exists
over an alphabet of size $q = 6m$, where $m$ is odd. In the next
step, we prove that an optimal code exists over an alphabet of
size $2q$ under the assumption that an optimal code exists over an
alphabet of size $q$.

We begin with the definition of a group divisible system. By an
$r$-subset of a set $X$, we mean a subset of $X$ with $r$ elements.
\begin{definition}[\cite{M}]
Let $m$ and $r$ be positive integers. Let $\mathcal{T}=\{T_1, T_2,
\ldots, T_m \}$ be a collection of disjoint $r$-sets whose union
is $T$. An $(m,r,k,b)$ group divisible system or a $G(m,r,k,b)$
system on $\mathcal{T}$ is a collection $\mathcal{B} = \{K_1, K_2,
\ldots, K_u \}$ of $k$-subsets of $T$ such that every b-subset in
$T$ is either contained in $T_i$ for some $i\in \bar{B}_{m}$ or it
is contained in a unique $k$-subset in $\mathcal{B}$ but not both.
\end{definition}
By a simple counting argument,
\begin{equation*}
|G(m,r,k,b)| = \frac{{mr \choose b} - m {r \choose b}}{{k \choose
b}}.
\end{equation*}
The following result was proved by Mills \cite{M}.
\begin{theorem}\label{Gm643}
A $G(m,6,4,3)$ system exists for every positive integer $m$.
\end{theorem}
Because we will use a subfamily $\mathcal{A}_1$ of a $G(m,6,4,3)$
system in our construction of an optimal code, we briefly review
the proof of Theorem \ref{Gm643} in \cite{M}, which introduces
$\mathcal{A}_1$ when $m$ is odd. Let $m$ be odd. For $i\in
\bar{B}_{m}$ let $T_i$ be the set of ordered pairs $(i,\alpha)$
with $\alpha \in B_6$ and $T$ be the union of these $T_i$. We
partition the elements of $B_6$ into pairs in three ways:
\begin{eqnarray*}
\begin{cases}
P_1 = \{(0,3), (1,5), (2,4)\},\\
P_2 = \{(1,4), (2,0), (3,5)\},\\
P_3 = \{(2,5), (3,1), (4,0)\}.
\end{cases}
\end{eqnarray*}
For any ordered pair $(w,x)\ (w<x)$ in $\bar{B}_{m}$ and any
$\lambda\in \bar{B}_3$, we form the nine quadruples
\begin{equation*}
\left\{(w,\alpha), (w, \beta), (x, \gamma), (x, \delta)\right\},
\end{equation*}
where $(\alpha, \beta)\in P_{\lambda}$ and $(\gamma, \delta) \in
P_{\lambda}$. This gives a collection of $\frac{27m(m-1)}{2}$
quadruples, denoted by $\mathcal{A}$, which is contained in
$G(m,6,4,3)$. Among the quadruples in $\mathcal{A}$, choose those
with $\lambda=1$ and denote them by $\mathcal{A}_1$. Note that
$|\mathcal{A}_1| = 9 {m \choose 2}$.

\begin{theorem}\label{modd}
$N(4,q,1) = \frac{q^2(q+2)}{4}$ for $q=6m$ where $m$ is odd.
\end{theorem}
\begin{pf}
We retain the notation used in the preceding discussion throughout
this proof. Let $m$ be odd and consider $G(m,6,4,3)$ on
$\mathcal{T} = \{T_1, T_2, \ldots, T_m\}$. Recall that our goal is
to construct an optimal code in $B_q^4$ where $q=|T|=6m$. Define
an order on $T$ as follows:
\begin{eqnarray*}
\begin{cases}
(i, \alpha) < (j, \alpha)\ (i < j),\\
(i, \alpha) < (j, \beta)\ (\alpha < \beta).
\end{cases}
\end{eqnarray*}
To each quadruple $\{x,y,z,w \}$ in $G(m,6,4,3)$ with $x<y<z<w$, we
associate the word $(x,y,z,w)$. From now on, we consider the
quadruples in $G(m,6,4,3)$ as words of length 4 defined as above.

Let
\begin{equation*}
M=\left(\bigcup \limits_{x \in \mathcal{A}_1} \langle
x\rangle_{B_{q}^4} \right)\bigcup\left(\bigcup \limits_{x \in
G(m,6,4,3) \setminus \mathcal{A}_1} \langle
x\rangle_{A_{q}^4}\right).
\end{equation*}
Note that for any two distinct elements $a = (a_1,a_2,a_3,a_4)$
and $b = (b_1,b_2,b_3,b_4)$ of $\mathcal{A}_1$, if $|L(a) \cap
L(b)| = 2$, then either $\{a_1,a_2\} = \{b_1,b_2\}$ or
$\{a_3,a_4\} = \{b_3,b_4\}$. From the structure of
$\mathcal{A}_1$, the definition of $G(m,6,4,3)$, and Lemma
\ref{deletion1}, we deduce that $M$ is a code in $B_q^4$ that is
capable of correcting single deletions. Since $|\mathcal{A}_1| = 9
{m \choose 2}$ and $|G(m,6,4,3)| = \frac{6m(6m-1)(6m-2)}{24} -
5m$, it follows that
\begin{equation*}
|M| = 8|\mathcal{A}_1| + 6|G(m,6,4,3)\setminus\mathcal{A}_1| =
54m^3 - 18m^2-36m.
\end{equation*}
Since $N(4,6,1) = 72$ (cf. Table \ref{table3}) and
$\left|T_i\right|=6$ for each $i\in\bar{B}_m$, we can choose an
optimal code $C_i$ over $T_i$ of cardinality 72 that is capable of
correcting single deletions. Let
\begin{equation*}
C= \left(\bigcup \limits_{i=1}^{m} C_i\right)\bigcup M \bigcup
\{(a,a,b,b)\ |\ a \in T_i, b \in T_j, i \neq j \}.
\end{equation*}
It is easy to check that $C$ is a code in $B_q^{4}$ that is
capable of correcting single deletions. From a simple calculation
\begin{eqnarray*}
&|C|& = m |C_1| + |M| + 2 \times 36 {m \choose 2} \nonumber \\
& & = 72m + 54m^3 -
18m^2 - 36m +36m(m-1) \nonumber \\
& & = 54 m^3 + 18m^2 = \frac{q^2(q+2)}{4}.
\end{eqnarray*}

This proves the theorem.
\end{pf}

The following lemma, originally due to Reiss \cite{R}, constructs
the systems that are equivalent to the systems $P_{\alpha}(m)$.
Because we need some terminology used in \cite{KS} to construct
optimal codes, we borrow a sketch of the proof from \cite{KS}.
\begin{lemma}[Reiss]\label{reiss} The $n(2n-1)$ pairs of
$2n$ elements can be partitioned into $2n-1$ sets $S_1,\,S_2,\,
\ldots,\,S_{2n-1}$ such that each set contains $n$ disjoint pairs.
\end{lemma}
\begin{pf}
Let $l_{ij} = i+j-1\ (\text{mod} \ 2n-1)$ where $i\in
\bar{B}_{2n-1}$, $j\in \bar{B}_{2n-1}$, and $l_{ij}\in
\bar{B}_{2n-1}$, and define $l_{i,2n} = l_{ii}\ (i\in
\bar{B}_{2n-1})$. Take
\begin{equation*}
S_q = \{(i,j)\ |\ i < j \ \text{and}\ l_{ij}=q\}\ (q\in
\bar{B}_{2n-1}).
\end{equation*}
Then these sets $S_q\ (q\in \bar{B}_{2n-1})$ have the desired
property.
\end{pf}

\begin{theorem}\label{meven}
If $N(4,q,1) = \frac{q^2(q+2)}{4}$ for $q = 6m$, then $N(4,2q,1) =
\frac{(2q)^2(2q+2)}{4}$.
\end{theorem}
\begin{pf}
Suppose that $N(4,q,1) = \frac{q^2(q+2)}{4}$ for $q = 6m$. Let
$\bar{B}_q(\alpha) = \{(a, \alpha)\ |\ a\in \bar{B}_q\}$
$(\alpha\in \bar{B}_2)$. We will construct a code in
$\bar{B}_q(1)\cup\bar{B}_q(2)$ that is capable of correcting
single deletions with cardinality $\frac{(2q)^2(2q+2)}{4}$. By
applying Lemma \ref{reiss} with $n=3m$, the pairs of elements of
$\bar{B}_q(\alpha)$ can be partitioned into the sets
$S_1^{\alpha},\,S_2^{\alpha},\,\ldots,\,S_{6m-1}^{\alpha}$, namely
\begin{equation*}
\text{the set of pairs of}\ \bar{B}_q(\alpha) = S_1^{\alpha} \cup
S_2^{\alpha} \cup \cdots \cup S_{6m-1}^{\alpha}\ (\alpha\in
\bar{B}_2),
\end{equation*}
where $S_l^{\alpha} = \left\{\left((a,\alpha), (b, \alpha)\right)\
|\ a < b, a+b-1 = l\ (\text{mod} \ 6m-1)\right\}$ for $\alpha\in
\bar{B}_2$ and $l\in \bar{B}_{6m-1}$. Note that $|S_l^{\alpha}| =
3m = \frac{q}{2}$ for all $\alpha$ and $l$. Consider the set
$S^{1,2}$ of ordered pairs defined as follows:
\begin{equation*}
S^{1,2}= \left\{(x, y)\ |\ x \in S_i^1, y \in S_i^2, i\in
\bar{B}_{6m-1}\right\} = \bigcup \limits_{i=1}^{6m-1}\left(S_i^1
\times S_i^2\right).
\end{equation*}
Since $x$ and $y$ are pairs of elements in $\bar{B}_q(\alpha)\
\left(\alpha\in \bar{B}_2\right)$, $S^{1,2}$ is a set of
quadruples of elements in $\bar{B}_q(1) \cup \bar{B}_q(2)$ whose
cardinality is $(6m-1)|S_1^1||S_1^2| = 9m^2(6m-1) =
\frac{q^2(q-1)}{2}$. Furthermore, $S^{1,2}$ has the following
property: every triple $\left\{(c_1,i),(c_2,j),(c_3,k)\right\}$
with elements from $\bar{B}_q(1)\cup\bar{B}_q(2)$, except triples
with element from only one of $\bar{B}_q(1)$ or $\bar{B}_q(2)$,
belongs to a unique quadruple in $S^{1,2}$.

We define an order on $\bar{B}_q(1) \cup \bar{B}_q(2)$ as follows:
\begin{eqnarray*}
\begin{cases}
(a, \alpha) < (b, \alpha) \ \mbox{if and only if} \ a < b,\\
(a, 1) < (b, 2) \ \mbox{for all} \ a, b.
\end{cases}
\end{eqnarray*}
To each quadruple $x = \{(a,1),(b,1), (c,2),(d,2) \}$ in $S^{1,2}$
where $((a,1),(b,1)) \in S_i^1$ and $((c,2),(d,2)) \in S_i^{2}$,
we associate a word $(a_1,a_2,a_3,a_4)$ of length 4 such that $a_1
< a_2 < a_3<a_4$ and $a_i \in \{(a,1),(b,1),(c,2),(d,2) \}$. From
now on, we consider the quadruples in $S^{1,2}$ as words of length
4 over $\bar{B}_q(1) \cup \bar{B}_q(2)$ defined as above. Let
\begin{equation*}
M= \left(\bigcup \limits_{x \in S_1^1 \times S_1^2} \langle
x\rangle_{B_{2q}^4}\right) \bigcup \left(\bigcup \limits_{x \in
S^{1,2} \setminus (S_1^1 \times S_1^2)} \langle
x\rangle_{A_{2q}^4}\right).
\end{equation*}

From the construction of $S^{1,2}$ and Lemma \ref{deletion1}, $M$
is a code in $\bar{B}_q(1) \cup \bar{B}_q(2)$ that is capable of
correcting single deletions. Note that
$$\left|M\right|=
8\left|S_1^1\right|\left|S_1^2\right|+6\left(\left|S^{1,2}\right|
-\left|S_1^1\right|\left|S_1^2\right|\right)=\frac{2q^2(3q-2)}{4}.$$
Since $N(4,q,1) = \frac{q^2(q+2)}{4}$ for $q = 6m$, there exists a
code $C^{\alpha}$ in $\bar{B}_q(\alpha)\ (\alpha\in \bar{B}_2)$
that is capable of correcting single deletions with cardinality
$|C^{\alpha}| = N(4,q,1) = \frac{q^2(q+2)}{4}\ (\alpha\in B_2)$.

Finally, let
\begin{equation*}
C= C^1 \cup C^{2} \cup M \cup \{(a,a,b,b),\,(b,b,a,a)\ |\ a \in
\bar{B}_q(1), b \in \bar{B}_q(2)\}.
\end{equation*}
It is easy to check that $C$ is a code in $\bar{B}_q(1) \cup
\bar{B}_q(2)$ that is capable of correcting single deletions with
$|C| = \frac{(2q)^2(2q+2)}{4}$. Because $\left|\bar{B}_q(1) \cup
\bar{B}_q(2)\right|=\left|B_{2q}\right|$ and $\bar{B}_q(1)\cap
\bar{B}_q(2)=\emptyset$, this proves the theorem.
\end{pf}

Combining Theorem \ref{modd} and \ref{meven} yields the following
theorem.

\begin{theorem}\label{6m}
There exists a code $C$ in $B_q^{4}$ that is capable of correcting
single deletions with $|C| =N(4,q,1)=\frac{q^{2}(q+2)}{4}$ for $q =
6m\ (m\ge 1)$.
\end{theorem}

Combining Theorems \ref{Steiner} and \ref{6m}, we finally obtain
the following.
\begin{theorem}\label{qeven}
For any even $q$, $N(4,q,1)=\frac{q^2(q+2)}{4}$.
\end{theorem}

\section{Perfect codes}

In this section, we modify our construction of optimal codes
slightly, and construct an optimal perfect code in $B_q^4$ when
$q$ is even.

We start with simple definitions. Recall that a code $C$ in
$B_q^n$ is an $s$-covering of $B_q^n$ from below if $\lfloor
C\rfloor_s=\left\lfloor B_q^n\right\rfloor_s$ \cite{L}. A code $C$
in $B_q^n$ that is capable of correcting $s$ deletions is called a
perfect code that is capable of correcting $s$ deletions if $C$ is
an $s$-covering of $B_q^n$ from below. For brevity a perfect code
in $B_q^{n}$ that is capable of correcting single deletions will
be referred to as a perfect code.

Levenshtein \cite{L} showed that there exists a perfect code in
$B_q^{4}$ of cardinality $\frac{q^3+q^2+2q}{4}$ for any even $q$.
Note that this code is not optimal. In previous sections, we have
constructed an optimal code $C$ in $B_q^{4}$ for any even $q$. By
counting the cardinality of $\left\lfloor C\right\rfloor_1$, one
can show that it cannot be a perfect code (for example, the
optimal code in Table \ref{table2} is not perfect). However, the
construction of an optimal code can be modified to obtain an
optimal perfect code as follows.

Suppose that $C$ is an optimal code in $B_q^{4}$ for an even $q$
which is constructed using the method in Section \ref{bound}.
Decompose $C$ into subcodes $C_1$, $C_2$, and $C_4$, where
$C_i=\{x\in C \bigm | |\lfloor x \rfloor_1 |=i\}$ for
$i\in\bar{B}_2$ or $i=4$. From the structure of $C$, the codes
$C_1$ and $C_2$ are as follows:
$$\left\{\begin{array}{l}
C_1=\{(a,a,a,a)\ |\ a \in B_q\},\\
C_2=\{(a,a,b,b)\ |\ \{a,\,b\}\subset B_q\ \text{and}\ a\neq b\}.
\end{array}\right.$$
Note that $|C_1| = q, |C_2| = q(q-1)$, and that $|C_4| =
\frac{q^2(q+2)}{4} - \left(q + q(q-1)\right)$. Now we modify the
subcode $C_2$ to make $C_2'$ as follows:
\begin{align*}
C_2' &= \left(C_2 \setminus \left\{(2t,2t,2t+1,2t+1)\ |\ t\in
B_{\frac{q}{2}-1}\right\}\right) \\
&\cup \ \left\{(2t,2t+1,2t,2t+1)\ |\ t\in B_{\frac{q}{2}-1}\right\}.
\end{align*}

It can be easily verified that $C' = C_1 \cup C_2' \cup C_4$ is a
code that is capable of correcting single deletions with $|C| =
|C'|$. Note that $|C_2'| = |C_2|$ and $\left|C_2'\setminus
C_2\right| = \frac{q}{2}$. For each $x \in C'$, the following
relations arise after single deletions:
$$\left\{\begin{array}{l}
|\lfloor x
\rfloor_1 |= 1\ (x \in C_1),\\
|\lfloor x
\rfloor_1 |= 2\ (x \in C_2),\\
|\lfloor x
\rfloor_1 |= 4\ (x \in C_2' \setminus C_2),\\
|\lfloor x \rfloor_1 |=4\ (x \in C_4).
\end{array}\right.$$
Computing the cardinality of $\lfloor C' \rfloor_1$ yields:
$$\begin{array}{lll}
\left|\left\lfloor C' \right\rfloor_1\right|&=&\left|\left\lfloor
C_1 \right\rfloor_1\right|+\left|\left\lfloor C_2' \right\rfloor_1
\right|
+ \left|\left\lfloor C_4 \right\rfloor_1\right|\\
   &=& q + 2\left( q(q-1) - \frac{q}{2}\right) + 4 \frac{q}{2}
   + 4 \left( \frac{q^2(q+2)}{4} - q  - q(q-1)\right)\\
   &=& q^3.
\end{array}$$
Hence $C'$ is an optimal perfect code.

Therefore we obtain the following theorem.

\begin{theorem} For any even $q$,
we can construct an optimal perfect code in $B_q^4$ that is
capable of correcting single deletions.
\end{theorem}

{\bf Acknowledgement.} The authors would like to thank to Dr.
Vladimir Levedev for introducing this problem to them.

\end{document}